\documentclass{aa}

\usepackage{longtable}

\usepackage{psfig}
\usepackage{supertabular}

\begin{document}
% ====================================================================
%                       MACROS PERSO
% =====================================================================
%
% -----
% Abrev
% -----
\def\lfir{$L_{\rm FIR}$}
\def\mabs{M$_{\rm abs}$}
\def\etal{et al.}
\def\hii{H{\sc ii}}
\def\cbeta{$c_{\rm H\beta}$}
\def\av{A$_{\rm v}$}
\def\flam{$F_{\lambda}$}
\def\ilam{$I_{\lambda}$}
\def\teff{$T_{\mathrm{eff}}$}
\def\lg{$\log g$}
\def\feh{$\mathrm{[Fe/H]}$}
\def\mh{$\mathrm{[M/H]}$}
\def\ltsima{$\buildrel<\over\sim$}
\def\lsim{\lower.5ex\hbox{\ltsima}}

% -----
% Units
% -----
\def\micron{$\mu$m}
\def\kms{km s$^{-1}$}
\def\kmsmpc{km s$^{-1}$ Mpc$^{-1}$}
\def\cmc{cm$^{-3}$}
\def\erg{ergs s$^{-1}$ cm$^{-2}$ \AA$^{-1}$}
\def\ergs{ergs s$^{-1}$}
\def\ergscm{ergs s$^{-1}$ cm$^{-2}$}
\def\msun{\ifmmode M_{\odot} \else M$_{\odot}$\fi}
\def\zsun{\ifmmode Z_{\odot} \else Z$_{\odot}$\fi}
\def\lsun{\ifmmode L_{\odot} \else L$_{\odot}$\fi}

% -----
% Photometry etc.
% -----
\def\ubvetc{(UBV)$_J$\-(RI)$_C$\- JHKLL$^\prime$M}
\def\basel {{\it B}a{\it S}e{\it L}}
\def\was{CMT$_1$T$_2$}
% -----
% references
% -----
%
\def\aap{A\&A}
\def\aaps{A\&AS}
\def\aas{A\&AS}
\def\aj{AJ}
\def\apj{ApJ}
\def\apjl{ApJ}
\def\apjs{ApJS}
\def\mnras{MNRAS}
\def\pasp{PASP}
%
%%%%%%%%%%%%%%%%%%%%%%%%%%%%%%%%%%%%%%%%%%%%%%%%%%%%%%%%%%%%%%%%%%%%%%%%
%\begin{document}

\thesaurus{08(08.07.1;08.05.3;08.08.1;08.06.3)} 

\title{Database of Geneva stellar evolution tracks and isochrones for \ubvetc,
HST-WFPC2, Geneva and Washington photometric systems
%\thanks{}
}

\author{Thibault Lejeune\inst{1,2} \and Daniel Schaerer \inst{3}}

\offprints{D. Schaerer, schaerer@ast.obs-mip.fr}

 \institute{Observat\'orio Astron\'omico  da Universidade de  Coimbra,
 Santa Clara, P-3040 Coimbra,   Portugal \and Astronomisches  Institut
 der Universit\"at Basel, Venusstr.  7, CH-4102 Binningen, Switzerland
 \and Observatoire Midi-Pyr\'en\'ees, Laboratoire d'Astrophysique, UMR
 5572, 14, Av.  E. Belin, F-31400 Toulouse, France }

% \institute{Astronomisches Institut der Universit\"at Basel, Venusstr.
% 7,  CH-4102 Binningen, Switzerland  \and Observat\'orio Astron\'omico
% da Universidade  de Coimbra,  Santa  Clara, P-3040  Coimbra, Portugal
% \and Observatoire Midi-Pyr\'en\'ees, Laboratoire d'Astrophysique, UMR
% 5572, 14, Av.  E. Belin, F-31400 Toulouse, France }

\date{Received 28 july 2000 / Accepted 7 november 2000}

\titlerunning{Database of stellar models in different photometric systems}
%\authorrunning{}
\maketitle

%%%%%%%%%%%%%%%%%%%%%%%%%%%%%%%%%%%%%%%%%%%%%%%%%%%%%%%%%%%%%%%%%%%%%%%%
\begin{abstract} 
We have used an updated version of the empirically and semi-empirically
calibrated \basel\ library of synthetic stellar spectra of Lejeune
\etal\ (1997, 1998) and Westera \etal\ (1999) to calculate synthetic
photometry in the \ubvetc,  HST-WFPC2, Geneva, and  Washington systems
for the entire   set of non-rotating  Geneva stellar  evolution models
covering  masses from 0.4--0.8 to   120--150 \msun\ and  metallicities
$Z$=0.0004 (1/50 \zsun) to 0.1 (5 \zsun).
The results are provided  in a database\footnote{The full database will 
be available in electronic form at the CDS 
{\tt http://cdsweb.u-strasbg.fr/cgi-bin/qcat?J/A+A/(vol)/(page)}
and at {\tt http://webast.ast.obs-mip.fr/stellar/}.}
which includes all individual
stellar tracks and   the corresponding isochrones covering ages   from
10$^3$ yr to 16--20 Gyr in time steps of $\Delta \log t=$ 0.05 dex.  The
database also includes a new grid of stellar tracks of very metal-poor
stars ($Z$=0.0004)  from 0.8 --  150 \msun\ calculated with the Geneva
stellar evolution code.


\end{abstract}

%%%%%%%%%%%%%%%%%%%%%%%%%%%%%%%%%%%%%%%%%%%%%%%%%%%%%%%%%%%%%%%%%%%%%%%%
\section{Introduction}
Studies   of individual  stars  and resolved   stellar populations are
fundamental  for a wide  variety  of astrophysical subjects.  In  most
cases photometric observations, obtained in various systems, represent
the main observable. 
To translate these observables into fundamental stellar parameters 
(mainly effective  temperature, luminosity) one first has to rely on 
stellar atmosphere models.
In a second step the observations can then be compared to stellar evolution 
models and interpreted in physical terms such as mass, age, composition etc.
Only in rare cases (very  low mass stars,  Baraffe \etal\ 1995; O stars, 
Schaerer \etal\ 1996) stellar evolution and atmosphere models are coupled
and predict thus directly observables such as colors, magnitudes etc.

Extensive  grids    of stellar tracks  covering    the  most important
evolutionary phases   and   a large  metallicity   range  have  become
available  in the nineties  (see e.g.\ compilation in Leitherer \etal\
1996).  In  addition, recent efforts have   been undertaken to provide
accurate synthetic  colours  from grids of  atmosphere models covering
the bulk of the parameter space occupied by observed stars of spectral
types from O  to M and  all luminosity  classes (Lejeune \etal\  1997,
1998, Bessell \etal\ 1998).
With   the availability  of   such data it    now becomes  feasible to
systematically convert stellar    tracks  of all stellar  masses   and
derived isochrones  to a variety  of photometric  systems.  To provide
such a tool is the main goal of our database.

Existing grids, partly fulfilling this aim, include the library of Padova
isochrones with UBVRIJHK photometry for ages $\sim$ 4 Myr -- 20 Gyr and 
metallicities $Z$ between 0.0004 and 0.05 (Bertelli \etal\ 1994, Girardi 
\etal\ 1996), the recent $\alpha$-element enhanced tracks and isochrones
of Salasnich \etal\ (2000) with UBVRIJHK and HST-WFPC2 photometry,
and various other calculations covering smaller fractions of the parameter
space.

For the present work we rely on an updated version of the hybrid
library of synthetic stellar spectra compiled by Lejeune \etal\ (1997,
1998), which is corrected to match empirical colour--temperature
relations at solar metallicity and semi-empirically corrected at other
metallicities. These atmosphere models are used to derive synthetic
photometry of the \ubvetc, WFPC2, Geneva, and Washington systems,
using proper reference spectra for the zero points and up-to-date
filter curves.
This procedure is applied to essentially the full set of Geneva stellar 
evolution models to yield the photometric data for these systems for
{\em 1)} individual stellar tracks in the mass range from 0.4--0.8
to 120--150 \msun\
and metallicities $Z$=0.0004 (1/50 \zsun) to 0.1 (5 \zsun), and
{\em 2)} corresponding isochrones covering ages from $\sim$ 0 to 16--20 Gyr.

The input tracks, atmosphere models,  and the synthetic photometry are
described in Sect.\ 2.  The  resulting database products  (tracks  and
isochrones) are presented in  Sect.\ 3. Concluding  remarks on the use
of the  various database sets  are given in  Section  4.  The Appendix
includes the relevant data for a previously unpublished set of stellar
tracks at $Z$=0.0004.

%%%%%%%%%%%%%%%%%%%%%%%%%%%%%%%%%%%%%%%%%%%%%%%%%%%%%%%%%%%%%%%%%%%%%%%%
\section{Input}
\subsection{Stellar models}
The current compilation includes all grids of Geneva stellar evolution
models  published between 1992 and march  1999 in papers  I--VIII 
(see footnote of Table  \ref{tab_grids} for complete references)
and a new grid of  stellar models   for  very  low  metallicity (Z=1/50  
\zsun,  see Appendix). 
Pre-main sequence models are not included
(but cf.\ Bernasconi 1996 and paper VIII). 
The complete  set  including  the  references,   a  grid  number  used
subsequently throughout the paper, and  other main characteristics are
listed in Table \ref{tab_grids}  (see the original  papers for  a full
description).

\begin{table*}[htb]
 \caption[]{Summary of stellar grids  included  in the database.   The
 grid number  used throughout  the paper  (col.\ 1), database  ID (2),
 initial   chemical composition (3--5),  indications on  the mass loss
 prescription (6), the  covered mass range  (7), the reference for the
 tracks (8), and a brief description of the main characteristics (9) 
 are given.
\label{tab_grids}
}
\begin{tabular}{llrrrrlll}
\hline
\noalign{\smallskip}
Grid    & database & $Z$  &  $X$  &  $Y$  & $\dot{M}$ & mass range & paper$^{(*)}$ & description \\
\#      & ID          \\
(1)     & (2)      & (3)  & (4)   & (5)   & (6)       & (7)        & (8)           & (9)         \\
\noalign{\smallskip}
\hline
\noalign{\smallskip}
\multicolumn{4}{l}{Basic grids:} \\
1  & e & 0.0004 & 0.7584 & 0.2412 &  2$\times$std & 0.8 -- 150  \msun & this paper & basic grid \\
2  & c & 0.001 & 0.756 & 0.243 &  standard  & 0.8 -- 120 \msun & I   & basic grid \\
3  & c & 0.004 & 0.744 & 0.252 &  standard  & 0.8 -- 120 \msun & III & basic grid \\
4  & c & 0.008 & 0.728 & 0.264 &  standard  & 0.8 -- 120 \msun & II  & basic grid \\
5  & c & 0.020 & 0.680 & 0.300 &  standard  & 0.8 -- 120 \msun & I   & basic grid \\
6  & c & 0.040 & 0.620 & 0.340 &  standard  & 0.8 -- 120 \msun & IV  & basic grid  \\
7  & c & 0.100 & 0.420 & 0.480 &  standard  & 0.8 -- 60  \msun & VII & basic grid \\
\\
\multicolumn{4}{l}{Extended grids:} \\
8  & e & 0.001 & 0.756 & 0.243 & 2$\times$std: $M \ge 25 \msun$ & 0.8 -- 120 \msun & V, I    & high mass loss for massive stars \\
9  & e & 0.004 & 0.744 & 0.252 & 2$\times$std: $M \ge 20 \msun$ & 0.8 -- 120 \msun & V, III  & high mass loss for massive stars \\
10  & e & 0.008 & 0.728 & 0.264 & 2$\times$std: $M \ge 15 \msun$ & 0.8 -- 120 \msun & V, II  & high mass loss for massive stars \\
11 & e & 0.020 & 0.680 & 0.300 & 2$\times$std: $M \ge 15 \msun$ & 0.8 -- 120 \msun & V, I    & high mass loss for massive stars \\
12 & e & 0.040 & 0.620 & 0.340 & 2$\times$std: $M \ge 12 \msun$ & 0.8 -- 120 \msun & V, IV   & high mass loss for massive stars \\
13 & p & 0.001 & 0.756 & 0.243 &  standard  & 0.8 -- 1.7 \msun & VI   & including HB and EAGB \\
14 & p & 0.020 & 0.680 & 0.300 &  standard  & 0.8 -- 1.7 \msun & VI   & including HB and EAGB \\
15 & m & 0.001 & 0.756 & 0.243 &  standard  & 0.4 -- 1.0 \msun & VIII & MHD equation of state \\
16 & m & 0.020 & 0.700 & 0.280 &  standard  & 0.4 -- 1.0 \msun & VIII & MHD equation of state \\
17 &   & 0.020 & 0.680 & 0.300 &  standard  & 0.4 -- 1.0 \msun & VIII & MHD equation of state \\\\
\multicolumn{4}{l}{Alternate and combined model sets:} \\
   & o & 0.0004& 0.7584& 0.2412&2$\times$std& 0.8 -- 2.5 \msun & this paper   & no overshooting: 1.25 \msun \\
   & o & 0.001 & 0.756 & 0.243 &  standard  & 0.8 -- 2.5 \msun & I   & no overshooting: 1.25 \msun \\
   & o & 0.004 & 0.744 & 0.252 &  standard  & 0.8 -- 2.5 \msun & III & no overshooting: 1.25 \msun \\
   & o & 0.008 & 0.728 & 0.264 &  standard  & 0.8 -- 2.5 \msun & II  & no overshooting: 1.25 \msun \\
   & o & 0.020 & 0.680 & 0.300 &  standard  & 0.8 -- 2.5 \msun & I   & no overshooting: 1.25 \msun \\
   & o & 0.040 & 0.620 & 0.340 &  standard  & 0.8 -- 2.5 \msun & IV  & no overshooting: 1.25 \msun \\
   & l & 0.001 & 0.756 & 0.243 &  standard  & 0.4 -- 2.5 \msun &     & combination: grids 15, 13, 2 \\
   & l & 0.020 & 0.680 & 0.300 &  standard  & 0.4 -- 2.5 \msun &     & combination: grids 17, 14, 5 \\

\noalign{\smallskip}
\hline
\end{tabular}
\begin{tabbing}
 $^{(*)}$ \= IV\= : Schaerer et al. (1993)\ \ \ \ \ \= VIII\= \kill
 $^{(*)}$ \> I  \> : Schaller et al. (1992)     \> V   \> : Meynet et al. (1994)   \\
          \> II \> : Schaerer et al. (1993)     \> VI  \> : Charbonnel et al. (1996) \\
          \> III\> : Charbonnel et al. (1993)   \> VII \> : Mowlavi et al. (1998a) \\
          \> IV \> : Schaerer et al. (1993)     \> VIII\> : Charbonnel et al. (1999) \\
\end{tabbing}
\end{table*}

\begin{table*}[htb]
 \caption[]{Summary of isochrone  sets  included in the database.  The
 database ID  (col.\ 1), metallicity  (2), and  a brief description of
 the  main  characteristics (5)  are  indicated.  The  logarithmic age
 range (in years) covered by the isochrones is  given in columns 3 and
 4.  Isochrones are provided  for time steps  of $\Delta \log  t=0.05$
 for ages above 1 Myr ($\log t=6.0$).  For younger ages isochrones are
 for $\log t=$ 3.0, 5.0, 5.3, 5.6, 5.8, and 5.9 yr.
\label{tab_iso}
}
\begin{tabular}{lrrrlll}
\hline
\noalign{\smallskip}
Database & $Z$  &  \multicolumn{2}{c}{age range ($\log t$)} & description \\
ID       &      &  from& to                                 &     \\
(1)      & (2)  & (3)  & (4)                                & (5) \\ 
\noalign{\smallskip}
\hline
\noalign{\smallskip}
\multicolumn{4}{l}{Basic grids:} \\
e & 0.0004& 3.00 & 10.20 & basic grid \\
c & 0.001 & 3.00 & 10.20 & basic grid  \\
c & 0.004 & 3.00 & 10.20 & basic grid  \\
c & 0.008 & 3.00 & 10.20 & basic grid  \\
c & 0.020 & 3.00 & 10.20 & basic grid  \\
c & 0.040 & 3.00 & 10.20 & basic grid  \\
c & 0.100 & 3.00 & 10.20 & basic grid  \\
\\
\multicolumn{4}{l}{Extended grids:} \\
e & 0.001 & 3.00 & 7.50  & high mass loss for massive stars  \\
e & 0.004 & 3.00 & 7.50  & high mass loss for massive stars  \\
e & 0.008 & 3.00 & 7.50  & high mass loss for massive stars  \\
e & 0.020 & 3.00 & 7.50  & high mass loss for massive stars  \\
e & 0.040 & 3.00 & 7.50  & high mass loss for massive stars  \\
p & 0.001 & 9.00 & 10.20 & including HB and EAGB phases for low mass stars \\
p & 0.020 & 9.00 & 10.20 & including HB and EAGB phases for low mass stars \\
m & 0.001 & 9.00 & 10.30 & MHD equation of state for low mass stars \\
m & 0.020 & 9.00 & 10.30 & MHD equation of state for low mass stars \\
\\
\multicolumn{4}{l}{Alternate and combined model sets:} \\
o & 0.0004& 9.00 & 10.20 & no overshooting: 1.25 \msun \\
o & 0.001 & 9.00 & 10.20 & no overshooting: 1.25 \msun \\
o & 0.004 & 9.00 & 10.20 & no overshooting: 1.25 \msun \\
o & 0.008 & 9.00 & 10.20 & no overshooting: 1.25 \msun \\
o & 0.020 & 9.00 & 10.20 & no overshooting: 1.25 \msun \\
o & 0.040 & 9.00 & 10.20 & no overshooting: 1.25 \msun \\
l & 0.001 & 9.00 & 10.30 & combination ``best'' low mass star models \\
l & 0.020 & 9.00 & 10.30 & combination ``best'' low mass star models \\
\noalign{\smallskip}
\hline
\end{tabular}
\end{table*}

\begin{table*}[htb]
\caption{Description of file extensions {\tt ext} used for the stellar tracks 
({\tt mod*.ext}) and isochrones ({\tt iso*.ext}).}
\begin{tabular}{lll}
\hline
\noalign{\smallskip}
File extension  & description \\
\noalign{\smallskip}
\hline
\noalign{\smallskip}
{\tt dat}        & Complete set of predicted surface stellar properties \\
{\tt UBVRIJHKLM} & Main stellar properties and photometric data for \ubvetc\ system \\
{\tt WFPC2}      & Main stellar properties and photometric data for WFPC2 system \\
{\tt geneva}     & Main stellar properties and photometric data for Geneva system \\
\noalign{\smallskip}
\hline
\end{tabular}
\label{ta_ext}
\end{table*}

The  physical ingredients have been   discussed in papers I--VIII. For
completeness sake  a brief summary of the  basic assumptions common to
most model sets is presented here.  Variations are discussed below.
{\em  1)}  OPAL and  low  temperature opacities  of   Kurucz (1991) or
Alexander \& Ferguson (1994) are used.
{\em  2)} The  initial $(Y,Z)$ composition  is derived  from  a linear
chemical enrichment law $Y=Y_P + (\Delta Y/\Delta Z) Z$ with $Y_P=0.24$, and
$\Delta Y/\Delta Z=$3.  for $Z\leq$0.02
and $\Delta Y/\Delta Z=$2.5 for $Z>0.02$ respectively.
{\em  3)}   Mass loss rates are    taken from de   Jager \etal\ (1998)
throughout  the HR-diagram except  on  the red  giant branch (RGB) and
early  asymptotic  giant branch  (EAGB)   for initial  masses $M  <  5 \msun$, 
where the following expression is used (cf.\ Reimers 1975):
 $\dot{M} = 4. \times 10^{-13} \eta L R / M$ in units of \msun\ yr$^{-1}$, 
with $\eta=0.5$ at solar metallicity (see Maeder \& Meynet 1989).
Mass loss  is  scaled with metallicity  by  $(Z/\zsun)^{0.5}$, 
except for Wolf-Rayet (WR) stars.
In the WR phases the relation of Langer (1989) for  WNE and WC stars,  
and $\dot{M}=4. \times  10^{-5}$ \msun\ yr$^{-1}$ for WNL stars is used.
{\em 4)} Moderate    core overshooting of $d_{\rm  over}/H_P=0.2$   is
included for stars $M \ge 1.5 \msun$. For stars at $M=1.25 \msun$ with
a small or absent convective core tracks with and without overshooting
are provided.
{\em 5)} 
In addition to the effects treated by Maeder \& Meynet (1989), partial 
ionisation  of   heavy elements  is included in  the equation of state.
{\em  6)} Optically thick  envelopes  of WR  stars are treated  in the
framework of the modified Castor, Abbott \& Klein (CAK, 1975) theory.

The above ingredients   are   used  in   grids 2-7  which  cover   the
metallicity range from $Z=0.001$ to 0.100 (1/20 --  5 \zsun).  Grids 1
and 8-12 were calculated with mass loss  rates enhanced by a factor of
2 during the MS, the pre-WR,  and WNL phases  for massive stars (range
indicated in  col.\ 6); for  lower masses the models  are complemented
with the tracks from grids 2-6 or  represent new calculations (grid 1;
see Appendix).
In grids 1-12,  depending on the stellar mass  range, the evolution is
in general followed  up to   the  following  phases:  to  the end   of
C-burning for massive stars ($M \ge 7 \msun$), to the end of the early
asymptotic   giant branch (EAGB) for intermediate   mass stars ($2 \le
M/\msun \le 5$), and up to before the helium flash
for low mass  stars  ($0.8 \le M/\msun  \le  1.7$).  Grids 13  and  14
(paper  VI) present  new  calculations for  the latter   mass range at
metallicities $Z=0.020$  and 0.001 including post-helium flash models,
i.e.\ covering   the horizontal  branch   (HB) and   EAGB phases.  The
post-helium    flash  tracks   do    not   include   overshooting  and
semi-convection.
The MHD (Mihalas, Hummer \& D\"appen, 1988) equation of state was used
for the low mass models of paper VIII (grids 15, 16, 17).  The
calculations in grids 15-17 are followed up to before the helium flash
or ages larger than the Hubble time ($\ga$ 20 Gyr).  Note that the
solar metallicity grid 16 uses a different $Y/Z$ composition than the
remaining $Z=0.020$ grids.

The present compilation includes stellar models  covering a very large
parameter  space  in terms  of   mass, metallicity,  and  evolutionary
phases.   However,  we wish to  stress again  that e.g.\ the following
evolutionary phases are not  covered: 1) pre-main sequence tracks (for
Geneva models see Bernasconi 1996 and paper VIII), 2) thermally pulsing
AGB stars, and   3)  post-AGB stars  and white  dwarfs.   Other phases
(e.g. horizontal branch) are given only for a subset of metallicities.
These  limitations should be recognized by   the user of the database.
Calculations from other   groups   partly including  such  phases  are
mentioned in Section 4.

%-----------------------------------------------------------------------
\subsection{Atmosphere models}

The conversion of the  theoretical tracks and isochrones of  the
present compilation  to observational colour-magnitude  (c-m) diagrams
has been  performed using the stellar  spectral  library of Lejeune
\etal\  (1997, 1998)  that  provides empirically  and semi-empirically
{\em colour-calibrated} model atmosphere  spectra for a large range of
fundamental stellar parameters,  $T_{\rm eff}$ (2000  K  to 50,000 K),
$\log g$ (-1.0 to 5.5), and {\feh} (5.0 to +1.0).

For the present calculations,  we use the  most recent version of this
library (Westera \etal\ 1999), {\basel}-2.2\footnote{The  {\basel}-1.0
and {\basel}-2.0 libraries were published in Lejeune
\etal\  (1997) and     Lejeune \etal\   (1998)  respectively,    while
{\basel}-2.2 is available only electronically at {\tt
ftp://ftp.astro.unibas.ch/pub/lejeune/}.  See also this ftp site for a
description of the changes in the different versions.}, in which (1)
all the model spectra of stars with {\teff} $\geq 10,000$ K are now
calibrated upon empirical colours from the $T_{\mathrm{eff}}$ {\it
versus} $(B-V)$ relation of Flower (1996), and (2) the calibration
procedure for the cool giant model spectra has been extended to the
parameter ranges 2500 K $\leq$ {\teff} $<$ 6000 K and -1.0 $\leq$
{\lg} $<$ 3.5
\footnote{In the  previous  versions   of the {\basel}
models,  we adopted  {\teff} =  5000  K and {\lg} =   2.5 as the upper
limits  for  the calibration of giants  (see  Lejeune \etal\  1998 for
details).},  with  in particular  the effect to  provide  redder model
colours for  giants,   in better agreement  with   observed  red giant
branches of metal-poor globular clusters (see Lejeune \& Buser 1999).

%-----------------------------------------------------------------------
\subsection{Synthetic photometry}
In   order   to     transform    the theoretical  quantities      ($L,
T_{\mathrm{eff}}$)  into magnitudes and  colours, we first compute, by
interpolation   in  the \basel\ grid      ({\teff}, {\lg}, and  {\feh}
\footnote{The solar metallicity \feh\  =0 is identified with the
$Z=0.02$ tracks.  For other metallicities, we adopt the relation
$\mathrm{[Fe/H]} \equiv \log(Z/Z_{\odot})$, with $Z_{\odot}= 0.02$.}
the model spectrum at each point along the track or the isochrone, and
then derive the synthetic photometry by convolving the resulting flux
distribution with the filter transmission functions (as given in the
following sections).

The  Wolf-Rayet phases  in the stellar   tracks are excluded from this
interpolation procedure because no  physical model is available in the
{\basel} grid to accurately describe the atmosphere and the spectra of
stars in these phases.  For these  reasons, in the  files we assign an
arbitrary     value  ($-99$)  to   the    magnitudes and  the  colours
corresponding to  these points.  It has  also to be pointed  out that,
because of  some missing  models in the   {\basel} set of models  near
{\teff} $\sim 11,000$ K and {\lg} $\sim  1.5$, extrapolations in $\log
g$  are  necessary to  compute  the colours  in this  parameter range,
leading  to a  few numerical  approximations.  As  a  consequence, the
derived    magnitudes and   colours     change  more  abruptly    with
$T_{\mathrm{eff}}$ than expected, and  hence the shape of the  stellar
tracks do  not appear very  smooth in this region  of the c-m diagram.
For these  peculiar  points, we  estimate  our computed photometry 
to be accurate to about $\pm 0.05$ mag.

\subsubsection{Absolute magnitudes and zero-points}
The absolute    magnitudes  are computed   from the   absolute
luminosity given in the tracks and the bolometric corrections computed
from stellar atmosphere models.  For {\em all} the photometric systems
described in the following  sections, we give the absolute  magnitude,
$M(V)$, expressed in  the Johnson band.    Absolute magnitudes in  the
other passbands  can then be derived  from the colour indices provided
in  the  files.   Our  adopted   bolometric  correction scale is  that
described in Lejeune \etal\  (1998), defined in  such a way to fit the
empirical  scale   of Flower  (1996)   instead   of using  the  unique
calibration point of the Sun.  Note that with this definition, we find
$BC(V)   =  -0.108$ for the solar   model  of  Kurucz  (1992), a value
slightly different of that usually quoted for the Sun.

The zero-points  of the   colours  are defined  from  the  Vega  model
spectrum  of  Kurucz (1992); for the   {\ubvetc}, the {\was},  and the
HST-WFPC2 photometric systems, the computed colours for this model are
set to zero, while for the Geneva system  these values are adjusted on
the     observed  photometry   of  Vega,  $U-B=1.505$,   $B1-B=0.959$,
$B2-B=0.900$,  $V1-B=1.510$, $V^{*}-B=1.662$, and $G-B=2.168$ (Rufener
1976), and   $V^{*}-V=0.0$\footnote{To  avoid the confusion   with the
Johnson $V$  band,  the  Geneva $V^{*}$   is noted with   an  asterisk
hereafter.}

\subsubsection{\ubvetc}
\label{ubvetc_photom}
For the Johnson-Cousins-Glass  photometry, we used the filter response
functions of Buser (1978) for $(UBV)$,  Bessell (1979) for $(RI)$, and
Bessell \& Brett (1988) for $JHKLL^{\prime}M$.

\subsubsection{WFPC2}
 HST-WFPC2  synthetic   photometry has  been  computed from  the
 passbands  described in the ``{\it WFPC2  Instrument Handbook}''.  We
 used the function responses\footnote{The tables can be retrieved from
 the              STScI         ftp         site                 ({\tt
 ftp://ftp.stsci.edu/cdbs/cdbs8/synphot{\_}tables/}) as  files    {\tt
 pcfxxxwsys.txt},  where    {\tt  xxx}  designates    the  mean filter
 wavelength in nm.} from May 1998 which include the effects of
 all the detection chain (filter transmission, CCD quantum efficiency,
 etc.), hence   allowing   direct comparisons of    our synthetic  c-m
 diagrams with HST  data.  The WFPC2  zero-points are usually  defined on
 the   $STMAG$     system,         based       on       a         {\em
 constant-flux-density-per-unit-wavelength}.   The conversion  between
 the $VEGAMAG$  system (based on the Vega's  spectrum) and the $STMAG$
 system  is given  by  the   following formula:
 $\mathrm{VEGAMAG} = \mathrm{STMAG} + 21.1 + 2.5
 *\log(F_{\nu}(\mathrm{Vega}))$ which, with our passband definitions
 and the Vega model spectrum of Kurucz, give the colour differences,
 $VEGAMAG-STMAG= -1.740$, $1.514$, $1.619$, and $1.292$ for
 F336W-F439W, F439W-F555W, F555W-F675W, and F675W-F814W respectively.

Colour-temperature and  bolometric corrections using  both the Lejeune
\etal\ (1997)  models and the Bessell \etal\  (1998) models have been
presented by  Origlia \& Leitherer (2000).   This paper  also includes
HST-NICMOS photometry  and information  on the  transformation between
WFPC2 and NICMOS and other systems.

\subsubsection{Geneva photometry}
The Geneva photometric indices, $U-B$, $B1-B$, $B2-B$, $V1-B$,
$V^{*}-B$, $G-B$, and $V^{*}-V$, have been computed using the latest
revision of the Geneva passband definitions (Nicolet, 1998, private
communication).
Differences with the passband functions previoulsy published by
Nicolet (1996) are very small, and do not exceed a few percents on the
computed colours. In order to test the accuracy of our computed
synthetic photometry, we made some comparisons with observed data for
a sample of dwarf stars with $\log g \sim 3.5$ to $5$ from the Rufener
(1988) catalogue.  The results have shown a good or very good
agreement with the observations for most of the colour indices, with
most of the colour diferences {\lsim} 0.05 mag. Larger deviations have
been found in the UV ($\sim$ 0.1 to 0.2 mag.  near {\teff} $= 8000$
K), and also for some peculiar multicolour parameters of the Geneva
system, such as $\Delta$ and $g$.

\subsubsection{Washington photometry}
 Washington broad-band photometry is often used in the studies of star
 clusters (see for instance Geisler \& Sarajedini 1999, Lejeune \&
 Buser 1999), and indeed provide very accurate photometric metallicity
 determinations for distant and extragalactic objects (Lejeune \&
 Buser 2001, in preparation).  We also give in the present database,
 colour transformations in the Washington system ($C, M, T_1, T_2$),
 using the filter response functions published by Canterna \& Harris
 (1979).  The Cousins $R$ band is very close to $T_1$ but has a higher
 efficiency, and hence could be a very good alternative to the
 Washington filter.  For this reason, we provide in the files some
 Washington colours based on the $R$ band instead $T_1$, as suggested
 by Geisler (private communication).

\subsubsection{Grids of colours}
Complete grids of  colours, computed from the whole {\basel}-2.2
library for the photometric systems described above are  available
via        anonymous      ftp         at        the       URL     {\tt
ftp://tangerine.astro.mat.uc.pt/pub/BaSeL/}.
 In the near future, an interactive on-line service allowing such
 colour conversions for arbitrary sets of stellar parameters will be
 available at the URL {\tt
 http://tangerine.astro.mat.uc.pt/BaSeL/}.

%%%%%%%%%%%%%%%%%%%%%%%%%%%%%%%%%%%%%%%%%%%%%%%%%%%%%%%%%%%%%%%%%%%%%%%%
\section{Database content and access}
The database includes all stellar  tracks listed in  column 2 of Table
\ref{tab_grids} with a database ID.  Corresponding isochrones covering
the age  range from 10$^3$  yr to 16 or 20  Gyr are also provided.  In
both cases the  data contains the  fundamental stellar parameters  and
all  predicted  surface  properties.  Files  including the   predicted
synthetic photometry are provided for  all tracks and isochrones.  The
organisation, nomenclature,  and content  of  the respective  files is
discussed in the following.

\subsection{Stellar tracks}
The   stellar tracks  are  grouped  in  files named  according to  the
following conventions: {\tt modIzzz.ext}, where {\tt I} stands for the
database  ID  (col.\  2, Table  \ref{tab_grids}), {\tt   zzz}  for the
metallicity, and  {\tt ext} for  the extension  specifying the type of
data (see    Table  \ref{ta_ext}).  For  example   {\tt modc001.WFPC2}
contains the WFPC2  photometric data  for all  stellar tracks of   the
model grid \# 2.
%(e.g.\ {\tt modc001.dat} for grid \# 2).

Grids 1-16 are included integrally in the database. The files with the
ID ``c'' or ``e'' contain the 1.25  \msun\ track calculated {\em with}
overshooting.  The corresponding {\em no}  overshoot model is found in
the ``o'' files, as indicated in col.\ 9 of Table \ref{tab_grids}.
The  ``l'' model set,  providing  the most  appropriate low mass  star
models, consist of the following  combination of tracks: 0.4--1 \msun\
main  sequence tracks from paper  VIII   (grids 15, 17  respectively),
post-He  flash models for 0.9  and 1 \msun\  from  paper VI (with ages
properly  adjusted  to the   calculations   of paper  VIII), 1.25  (no
overshoot), 1.5, and  1.7 \msun\  tracks from  paper VI  (grids 13, 14
respectively), and 2 and 2.5 \msun\ tracks from paper I (grids 2, 5).

Obviously some  of the  sets in the  database contain  redundant data.
This is due to the organisation of the sets  designed for a simple and
covenient use. See Section \ref{s_use} for some recommendations on how
to use the database.

\begin{figure*}[htbp]
\centerline{\psfig{figure=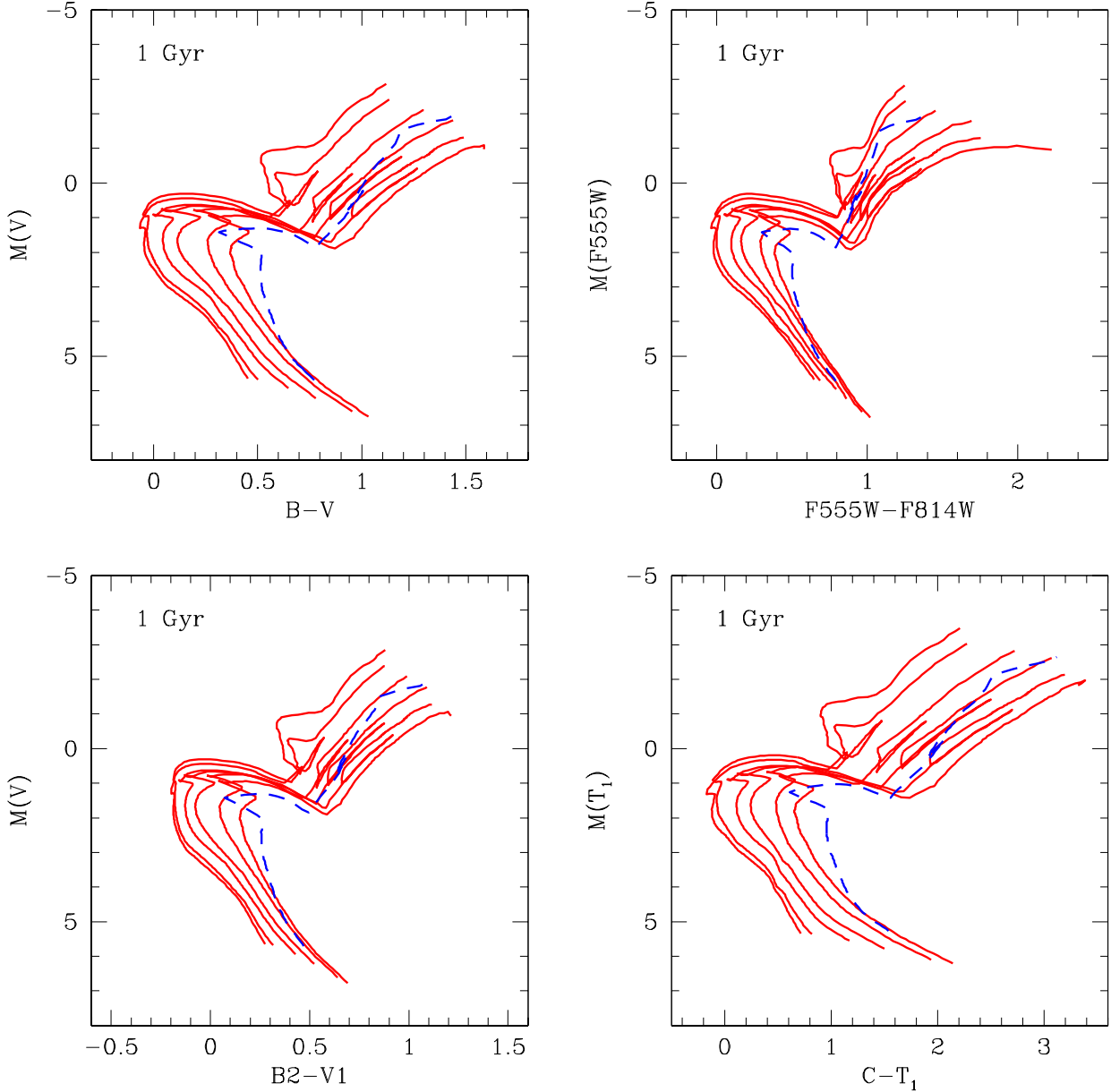,width=18cm}}
 \caption{Synthetic      colour-magnitude     diagrams  in   different
 observationnal  planes 
 (upper left: Johnson, upper right: HST-WFPC2, lower left: Geneva,
  lower right: Washington system) for the $1$ Gyr isochrones at different
 metallicities: from left to right in each panel, $Z =$ 0.0004, 0.001,
 0.004, 0.008, 0.02, 0.04, and $Z =$ 0.1 (dashed line).}
\label{plot_CMD}
\end{figure*}

The content of the various files is as follows:
\begin{itemize}
\item {\tt *.dat} files: 
 Predicted {\em surface}  stellar properties  using the table   format
 described in paper I (column 1-15),  complemented by the stellar core
 temperature for Wolf-Rayet  stars or  $T_{\rm eff}$ otherwise   (col.\
 16),\ and the mass loss rate ($\log  \dot{M}$ in \msun/yr; col.\ 17).
 Column 1-15 contain the   following values: ID of evolutionary  point
 (col.\   1), age (2), present  mass  (3), $\log L/L_\odot$ (4), $\log
 T_{\rm eff}$  (5), and the surface  abundances in mass fraction of H,
 $^4$He,  $^{12}$C, $^{13}$C, $^{14}$N,  $^{16}$O, $^{17}$O, $^{18}$O,
 $^{20}$Ne, and $^{22}$Ne (cols.\ 6-15).
 The {\em central} properties are given in the tables of the original 
 papers.

\item {\tt *.UBVRIJHKLM} files: 
  ID of evolutionary point (col.\ 1), age (2), present mass (3), $\log
  T_{\rm eff}$  (4), $\log g$ (5),  $\log  L/L_\odot$ (6), $M(V)$ (7),
  $U-B$ (8),   $B-V$ (9), $V-R$  (10),  $V-I$ (11), $V-K$  (12), $R-I$
  (13),  $I-K$ (14), $J-H$ (15), $H-K$  (16),  $K-L$ (17), $J-K$ (18),
  $J-L$ (19), $J-L2$ (20), $K-M$ (21).

\item {\tt *.WFPC2} files: 
  ID of evolutionary point (col.\ 1), age (2), present mass (3), $\log
  T_{\rm eff}$ (4), $\log g$ (5), $\log L/L_\odot$ (6), $M(V)$ (7),
  F336W-F439W (8), F439W-F555W (9), F555W-F675W (10), F555W-F814W
  (11), F675W-F814W (12), F439W-$V$ (13).

\item {\tt *.geneva} files: 
  ID of evolutionary point (col.\ 1), age (2), present mass (3), $\log
  T_{\rm eff}$  (4), $\log g$ (5),  $\log L/L_\odot$  (6), $M(V)$ (7),
  $U-B$ (8),  $B1-B$ (9),  $B2-B$ (10),  $V1-B$ (11),  $V^{*}-B$ (12),
  $G-B$ (13), $B1-B2$ (14), $B2-V1$ (15), $V1-G$ (16), $V^{*}-V$ (17).

\item {\tt *.CMT1T2} files: 
  ID of evolutionary point (col.\ 1), age (2), present mass (3), $\log
  T_{\rm eff}$ (4), $\log g$ (5), $\log L/L_\odot$ (6), $M(V)$ (7),
  $C-M$ (8), $M-T_1$ (9), $T_1-T_2$ (10), $C-T_1$ (11), $M-T_2$ (12),
  $C-R$ (13), $M-R$ (14), $R-T_2$ (15), $V-T_1$ (16).

\end{itemize}

\subsection{Isochrones}
For each of the database sets a large number of isochrones have been
calculated.  A summary of the available data, together with the
covered age range is shown in Table \ref{tab_iso}.  Isochrones are
provided for timesteps of $\Delta \log t=0.05$ for ages above 1 Myr
($\log t=6.0$). For younger ages isochrones are available for $\log
t=$ 3.0, 5.0, 5.3, 5.6, 5.8, and 5.9 yr.

The isochrones are grouped in files named according to the following
conventions: {\tt iso\_Izzz\_ffff\_tttt.ext}, where {\tt I} stands for
the database ID (col.\ 2), {\tt zzz} for the metallicity, {\tt ffff}
and {\tt tttt} indicate the logarithmic age range (from -- to) covered
by isochrones, and {\tt ext} for the extension specifying the type of
data (see Table \ref{ta_ext}).  For example {\tt
iso\_c001\_0300\_1020.UBVRIJHKLM} includes the \ubvetc\ photometry of
all isochrones with ages from 10$^3$ yr to 10$^{10.2}$ yr ($\sim$ 15.8
Gyr) of grid \# 2.
%(e.g.\ {\tt iso\_c001\_0300\_1020.dat} for grid \# 2).

The content of the files is the same as for the stellar tracks
described above, with the following exceptions: column 1 is replaced
by a line number, and column 2 now gives the initial mass along the
isochrone.

%%%%%%%%%%%%%%%%%%%%%%%%%%%%%%%%%%%%%%%%%%%%%%%%%%%%%%%%%%%%%%%%%%%%%%%%
\subsection{Illustrations}

In figure \ref{plot_CMD}, we give an illustration of some synthetic
c--m diagrams obtained from the present database.  The figure shows in
various planes the $1$ Gyr isochrones at the different metallicities
available ($Z =$ 0.0004, 0.001, 0.004, 0.008, 0.02, 0.04, and 0.1).
The peculiar behaviour of the $Z=0.1$ isochrone is due to the
overluminosity and the hotter temperature of these very metal-rich
stars discussed in detail in Mowlavi et al. (1998ab).

\subsection{Access to database}
The data will be accessible through the CDS\footnote{ 
{\tt http://cdsweb.u-strasbg.fr/cgi-bin/qcat?J/A+A/(vol)/(page)}}
and on the Web page {\tt http://webast.ast.obs-mip.fr/stellar/}.  
Future updates of the
database, including e.g.\ additional photometric systems, are foreseen
through the latter site.

%%%%%%%%%%%%%%%%%%%%%%%%%%%%%%%%%%%%%%%%%%%%%%%%%%%%%%%%%%%%%%%%%%%%%%%%
%\section{Discussion and conclusions}
\section{Recommendations for use}
\label{s_use}
We now briefly indicate the most appropriate set of models for the 
cases where several choices are offered.

For  most purposes the ``basic model  set'' (files  with the ID ``c'')
should be appropriate. The tracks of this set  have e.g.\ been applied
to analysis  of Galactic open clusters with  ages $\sim$ 4 Myr  -- 9.5
Gyr by Meynet \etal\ (1993).

For analysis  of massive  stars ($M_{\rm initial}  \ga  12 \msun$) the
high mass loss models (ID:  ``e'') should preferrentially be used (see
Maeder \& Meynet 1994).

For studies of low mass stars ($M_{\rm initial} \sim$ 0.4 -- 2.5
\msun) the combined model set (ID: ``l'') at Z=0.001 and Z=0.02 should
be most appropriate since it includes the MHD equation of state for $M
\le$ 1 \msun\ and covers also the post-Helium flash evolutionary phases
of stars with masses 0.8 $\le M/\msun \le$ 1.7.

For studies limited to the lowest  masses ($M \le$  1 \msun) the ``m''
set should be preferred.

For  studies critical  to the properties  of  stars in  the mass range
1. -- 1.5 \msun, corresponding to the appearance of core convection on
the main sequence,  the ``o'' sets allow  in comparison with the basic
sets  (``c'')  to study the   effect  of  enhanced  convective   cores
(overshooting) in this mass range.

As mentioned earlier,
the Geneva stellar grids used in this work do not include thermally
pulsing AGB stars and post-AGB phases, such as central stars of 
planetary nebulae (CSPN) and white dwarfs (WD).
If necessary the present data must be complemented with existing
calculations from the literature.
For recent calculations of AGB stars we refer e.g.\ to
the complete or combined complete + synthetic models of Vassiliadis \& 
Wood (1993), Bl\"ocker (1995a), Forestini \& Charbonnel (1997), 
and Langer \etal\ (1999), and to the synthetic models of 
Van den Hoeck \& Groenewegen (1997) and Marigo \etal\ (1998).
Recent models of CSPN and WD are given by e.g.\ Bl\"ocker (1995b)
Driebe \etal\ (1998), and Hansen \& Phinney (1998).

%%%%%%%%%%%%%%%%%%%%%%%%%%%%%%%%%%%%%%%%%%%%%%%%%%%%%%%%%%%%%%%%%%%%%%%%
\appendix
\section{Grid of stellar models at Z=0.0004}
A new grid of stellar models at very low  metallicity (1/50 solar) has
been  calculated by one  of us for  various applications (see de Mello
\etal\ 1998, Stasi\'nska  \& Schaerer 1999). The  full grid of  stellar
tracks from 0.8 to 150 \msun\ is briefly summarised here.

The  input physics is identical to  that of Meynet \etal\ (1994, paper
V).   This includes in particular  the adoption  of ``high'' mass loss
rates, i.e.\ $\dot{M}=8. \, 10^{-5}$ \msun/yr  for WNL stars and twice
the mass loss  rates of de Jager  \etal\ (1988) for stars with $M_{\rm
initial} \ge 3  \msun$. Otherwise we  follow the prescription of paper
I.  For consistence with the earlier  grids the initial composition is
calculated as in paper I: X=0.7584, Y=0.2412, Z=0.0004.

The    HRD with    the    calculated   tracks  is   shown   in   Fig.\
A1 %\ref{fig_hrd_0004}.   
Hatched  areas indicate  regions of slow nuclear
burning phases.  The H, He, and C burning lifetimes are given in Table
\ref{ta_lifetime}.  Due to numerical instabilities the 85 \msun\ model
could not  be fully evolved to  the  end of  He burning. The remaining
He-burning lifetime    as well  as the   duration   of C-burning  were
estimated by interpolation between the two adjacent tracks.
For similar  reasons $t_C$ of the  20 \msun\ model had also
to be estimated.

The mass  limit  for the  formation of  WR   stars, $M_{\rm WR}$,   at
Z=0.0004 from the  present tracks is  $M_{\rm  WR} \sim$ 85 \msun,  as
already derived  in de Mello \etal\ (1998).   Other properties  of the
Z=0.0004 models can readily be derived from the detailed data included
in the present database.

% % % % % % % % % % % % % % % % % % % % % % % % % % % % % % % % % % % % 
\begin{figure*}[htbp]
\centerline{\psfig{figure=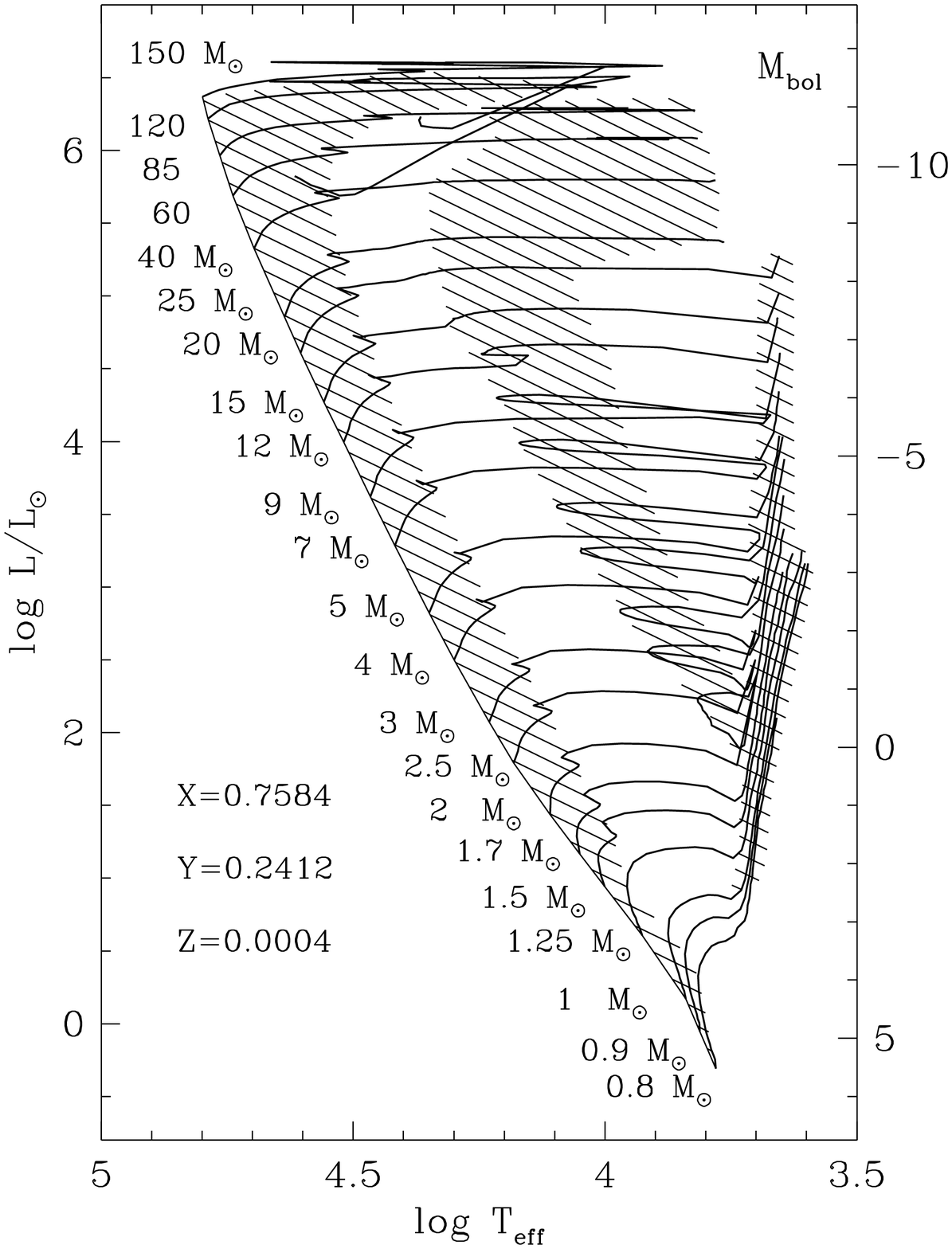,width=18cm}}
\label{fig_hrd_0004}
 \caption{Theoretical HR diagram for   the ensemble of the  calculated
 models   for metallicity   Z=0.0004  with an  overshooting  parameter
 $d/H_p=0.20$.   The slow phases  of nuclear  burning are indicated by
 hatched areas.  The $1.25 \rm M_\odot$ track corresponds to the model
 without overshooting}
\end{figure*}

\def\f{\frac}
\def\Msolar{$M_\odot$}
\def\h{\hphantom{9}}
\def\hq{\hphantom{9999}}
\def\hs{\hphantom{999}}
\def\hd{\hphantom{99}}
\def\tiret{\multicolumn{2}{c}{\mbox{---}}}

\begin{table}
\caption{\it Lifetimes in nuclear phases (in units of $10^6$ yr) for $Z = 0.0004$}
$${\begin{array}{r@{}l@{}lr@{.}lr@{.}lr@{.}lr@{.}l}  \hline \\[1mm] 
\multicolumn{3}{c}{\mbox{Initial}}& 
\multicolumn{2}{c}{\mbox{H-burning}}& 
\multicolumn{2}{c}{\mbox{He-burning}} &  
\multicolumn{2}{c}{\mbox{C-burning}} & 
\multicolumn{2}{c}{   \smash{ \displaystyle\f{t_{He}}{t_H} }  } \\ 
\multicolumn{3}{c}{\mbox{mass}}   & 
\multicolumn{2}{c}{\mbox{phase}} &  
\multicolumn{2}{c}{\mbox{phase}} &  
\multicolumn{2}{c}{\mbox{phase}} & 
\multicolumn{2}{c}{} \\[1mm]
 \hline
\\[0.5mm]
150 &\,{\rm M_\odot}& & 2&7239 & 0&2882 & 0&0056 & 0&1058 \\
120 &       & & 2&8937 & 0&3183 & 0&0053 & 0&1100 \\
 85 &       & & 3&3192 & 0&3033^{1} & 0&004^{1} & 0&0882 \\
 60 &       & & 3&9456 & 0&3359 & 0&0036 & 0&0851 \\
 40 &       & & 5&0948 & 0&4169 & 0&0055 & 0&0818 \\
 25 &       & & 7&5571 & 0&6235 & 0&0099 & 0&0825 \\
 20 &       & & 9&5360 & 0&8028 & 0&015^{1} & 0&0842 \\
 15 &       & & 13&4686 & 1&1523 & 0&0240 & 0&0856 \\
 12 &       & & 18&3778 & 1&5782 & 0&0436 & 0&0859 \\
  9 &       & & 29&0134 & 2&5926 & 0&0909 & 0&0894 \\
  7 &       & & 45&2717 & 4&4674      & \tiret & 0&0987 \\  
  5 &       & & 86&8714 & 10&1487     & \tiret & 0&1168 \\
  4 &       & & 137&3413 & 18&8569    & \tiret & 0&1373 \\
  3 &       & & 259&7664 & 41&2311    & \tiret & 0&1587 \\
  2 &.5     & & 399&8616 & 75&6684    & \tiret & 0&1892 \\
  2 &       & & 706&4008 & 136&2883   & \tiret & 0&1929 \\
  1 &.7     & & 1110&0520 &  \tiret & \tiret & \tiret \\
  1 &.5     & & 1603&4858 &  \tiret & \tiret & \tiret \\
  1 &.25 & \alpha = 0.2 & 2827&6495 & \tiret & \tiret &  \tiret \\
  1 &.25 & \alpha = 0.0 & 2623&3869 & \tiret & \tiret &  \tiret \\
  1 &       & &  6008&6523 & \tiret & \tiret &  \tiret \\
  0 &.9     & &  8963&7673 & \tiret & \tiret &  \tiret \\
  0 &.8     & & 14095&3710 & \tiret & \tiret &  \tiret \\
 \hline \\[-2mm]
  ^{1} & \multicolumn{10}{l}{{\mbox{estimated (cf. text)}}} \\[1mm]
\hline
\end{array}} $$
\label{ta_lifetime}
\end{table}

%%%%%%%%%%%%%%%%%%%%%%%%%%%%%%%%%%%%%%%%%%%%%%%%%%%%%%%%%%%%%%%%%%%%%%%%
\begin{acknowledgements}
We thank  Gilbert  Burki, No\"el  Cramer and Bernard  Nicolet for help
with the Geneva   photometric system.  We  have benefited  from useful
discussions  with  Roland Buser,  Corinne  Charbonnel, Andr\'e Maeder,
Jean-Claude Mermilliod,  and Georges Meynet.     Eva Grebel and  Eline
Tolstoy   provided   some early user  feedback   on  the database.  TL
gratefully  acknowledges  financial support   from the Swiss  National
Science  Foundation (grant 20-53660.98 to Prof.   Buser)  and from the
``Funda{\c{c}}\~ao para a  Ci\^encia e Tecnologia'' (Portugal), (grant
PRAXIS-XXI$/$BPD$/22061/99$).

\end{acknowledgements}

%%%%%%%%%%%%%%%%%%%%%%%%%%%%%%%%%%%%%%%%%%%%%%%%%%%%%%%%%%%%%%%%%%%%%%%%

%%%%%%%%%%%%%%%%%%%%%%%%%%%%%%%%%%%%%%%%%%%%%%%%%%%%%%%%%%%%%%%%%%%%%%%%
\end{document}